\documentclass[a4paper]{article}
\usepackage[utf8]{inputenc}
\usepackage[T1]{fontenc}

\usepackage[a4paper,left=2cm]{geometry}

\usepackage[section]{placeins}  
\usepackage{subfig}
\usepackage[english]{babel}

\usepackage{mathtools}

\usepackage[retain-explicit-plus]{siunitx}
\usepackage{interval}

\usepackage[]{cleveref}  
\crefname{figure}{Fig.}{Figs.}  
\crefname{equation}{equations}{Equations}

\usepackage{csquotes}  
\usepackage[style=nature,isbn=false,url=false,date=year,backend=biber]{biblatex}
\usepackage{authblk}

\usepackage{setspace, caption}
\usepackage{xspace}  
\newcommand\SiN{\ensuremath{\text{Si}_3\text{N}_4}\xspace}

\newcommand\SIparam[3]{\ensuremath{#1=\text{\SI{#2}{#3}}}}

\title{In-situ single-shot diffractive fluence mapping for X-ray free-electron laser pulses}

\author[1]{Michael Schneider}
\author[2]{Christian M. Günther}
\author[1]{Bastian Pfau}
\author[3]{Flavio Capotondi}
\author[3]{Michele Manfredda}
\author[3,4]{Marco Zangrando}
\author[3]{Nicola Mahne}
\author[3]{Lorenzo Raimondi}
\author[3]{Emanuele Pedersoli}
\author[1,2]{Stefan Eisebitt}

\affil[1]{Max-Born-Institut Berlin, 12489 Berlin, Germany}
\affil[2]{Institut für Optik und Atomare Physik, Technische Universität Berlin, 10623 Berlin, Germany}
\affil[3]{Elettra Sincrotrone Trieste, Trieste 34149, Italy}
\affil[4]{Istituto Officina dei Materiali, Consiglio Nazionale delle Ricerche, 34149 Trieste, Italy}

\date{\today}

\begin{document}
\maketitle
\section*{}
\textbf{
Free-electron lasers (FEL) in the extreme ultraviolet (XUV) and X-ray regime opened up the possibility for experiments at high power densities, in particular allowing for fluence-dependent absorption and scattering experiments to reveal non-linear light--matter interactions at ever shorter wavelengths.
Findings of such non-linear effects in the XUV and X-ray regime are met with tremendous interest, but prove difficult to understand and model due to the inherent shot-to-shot fluctuations in photon intensity and the often structured, non-Gaussian spatial intensity profile of a focused FEL beam.
Presently, the focused beam spot is characterized and optimized separately from the actual experiment. 
Here, we present the first simultaneous measurement of diffraction signals from solid samples in tandem with the corresponding single-shot spatial fluence distribution on the actual sample.
This new \textit{in-situ} characterization scheme enables fast and direct monitoring and thus control of the sample illumination which ultimately is necessary for a quantitative understanding of non-linear light--matter interaction in X-ray and XUV FEL experiments.
}
\section*{}
The development of free-electron lasers for the XUV and X-ray regime has been one of the major leaps in photon-based science in the last few decades.
It enabled key advances in the study of ultrafast dynamics of excitations in matter with a unique combination of coherent femtosecond pulses, ultrahigh intensities up to several \si{J/cm^2} and short wavelengths down to single nanometers and below \cite{Feldhaus2005}.
In recent years, observations of non-linear effects in solids such as four-wave mixing \cite{Bencivenga2015, Glover2012} and stimulated emission \cite{Beye2013,Stohr2015,Nagler2009} have been reported and underline the fact that a precise knowledge of the exact number of photons per unit time and area on the sample is crucial to interpret and devise meaningful experiments in this field.

A number of well-established techniques exist to estimate this pivotal parameter.
Gas monitor detectors (GMD) are able to measure the total photon number in a single, few-femtosecond pulse \cite{Tiedtke2008}, but cannot account for the intensity distribution \emph{within} the focal spot on the sample.
This distribution is typically measured -- separately from the actual experiment -- using wave-front sensing \cite{Keitel2016,LePape2002,Schafer2002}, ablative imprints \cite{Chalupsky2010,Nelson2009}, or by detecting the transmitted intensity through a small aperture scanned across the beam in the sample plane.
These approaches are highly invasive and cannot be performed in tandem with the majority of FEL experiments.
They are in particular incompatible with all scattering experiments in the forward direction and cannot account for the finite acceptance of a sample smaller than the beam size or for the beam position on a larger and potentially inhomogeneous sample.
This leads to significant uncertainties, especially in ``diffract-and-destroy'' experiments, where a new sample is aligned after every single shot \cite{Vartanyants2011,Chapman2006,Seibert2011}.

\begin{figure}
    \centering
    \includegraphics[width=170mm]{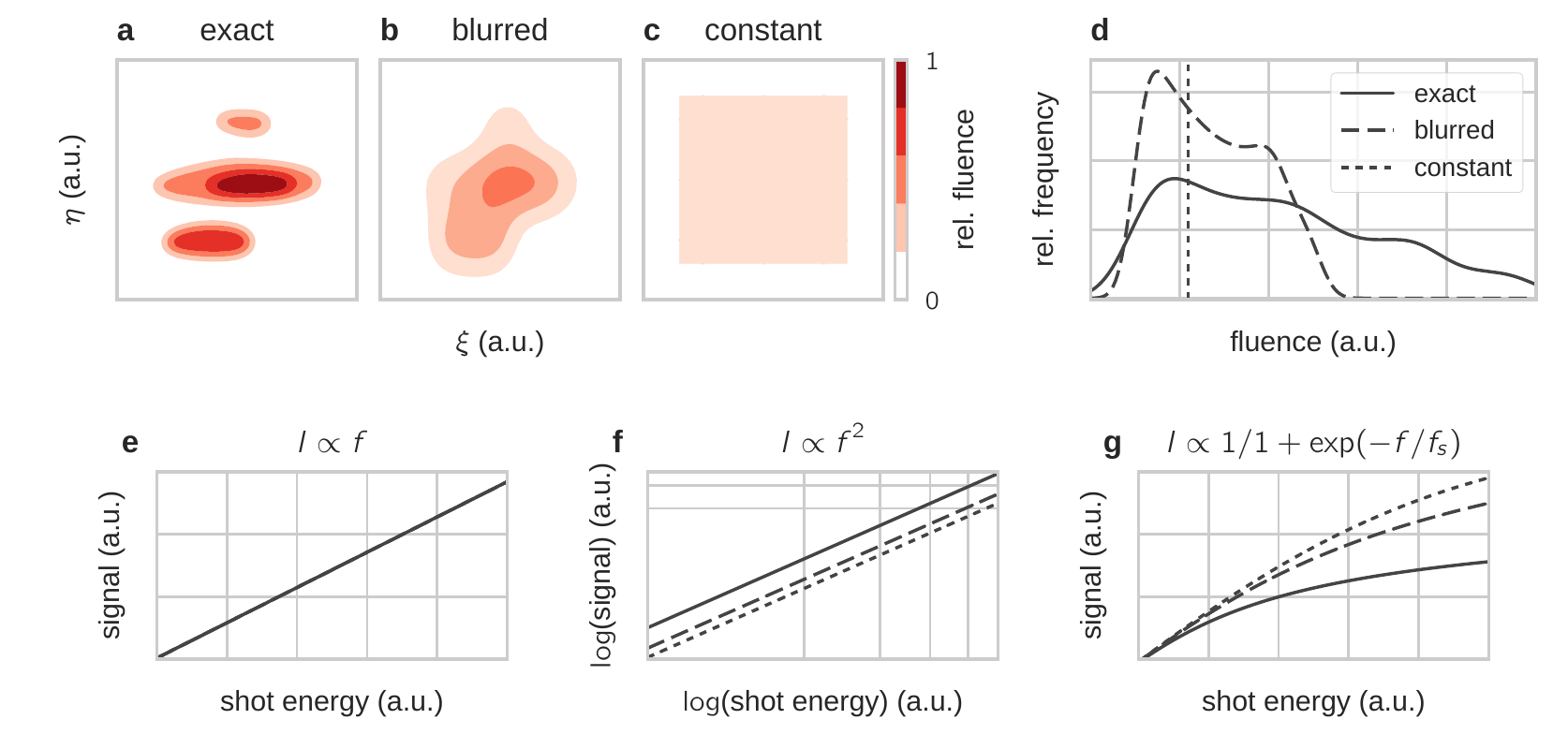}
    \caption{\textbf{Examples of spatial fluence distribution estimates.}
    Two-dimensional spatial fluence maps in the sample plane $(\xi, \eta)$, representing \textbf{a)} an exact measurement, \textbf{b)} a blurred, low-resolution measurement and \textbf{c)} a constant estimate.
    All maps sum to the same overall shot energy.
    \textbf{d)} Plot of the relative frequency of fluences $f$ in the three maps.
    \textbf{e)--g)} Simulated scattering signal $I(f)$ for increasing overall shot energy in the three fluence distributions under the assumption of, respectively, linear, quadratic and saturating fluence dependency.
    Note the double logarithmic scale in \textbf{f)}.
    }
    \label{fig:error_schematic}
\end{figure}

For demonstration purposes, in \cref{fig:error_schematic}, we schematically show how such uncertainties affect the interpretation of signals with a non-linear fluence dependency.
We consider three different fluence distribution estimates of a strongly focused FEL beam with a complex, non-Gaussian beam.
The distributions represent, respectively, an exact measurement (\cref{fig:error_schematic}a), a ``blurred'', low-resolution estimate as is typically the result of an aperture scan (\cref{fig:error_schematic}b) and a constant estimate, i.e.\ the characterization by a single ``spot size'' parameter (\cref{fig:error_schematic}c).
The fluence histograms vary drastically for the different estimates, as shown in \cref{fig:error_schematic}d.
For each fluence distribution, we calculate the signal levels assuming a linear, power-law or saturating fluence dependency (\cref{fig:error_schematic}e--g, respectively).
It is evident that, except for the linear case, an inaccurate fluence distribution obfuscates some or all of the characteristic parameters of the effect under study.

Thus, a correct interpretation of fluence dependent measurements will only be possible with an accurate, \textit{in-situ} characterization of the incident photon distribution $I(\xi, \eta)$ on the sample on a shot-by-shot basis.
In this Letter, we demonstrate such a simultaneous measurement of the spatial fluence distribution on the sample in conjunction with a diffraction signal from a solid sample in a single-shot XUV FEL experiment.
Our measurement scheme is derived from the work on monolithically integrated gratings on carrier membranes \cite{Schneider2016} and from the theoretical treatment of zone-plate diffraction under off-axis illumination \cite{Janicijevic1987}.
Via our integrated diffraction monitor design, we are able to map the incident photon distribution on the sample to the detector plane.
There, the illumination $I(x, y)$ is recorded simultaneously with the sample's scattering signal.

\begin{figure}
    \centering
    \includegraphics[width=170mm]{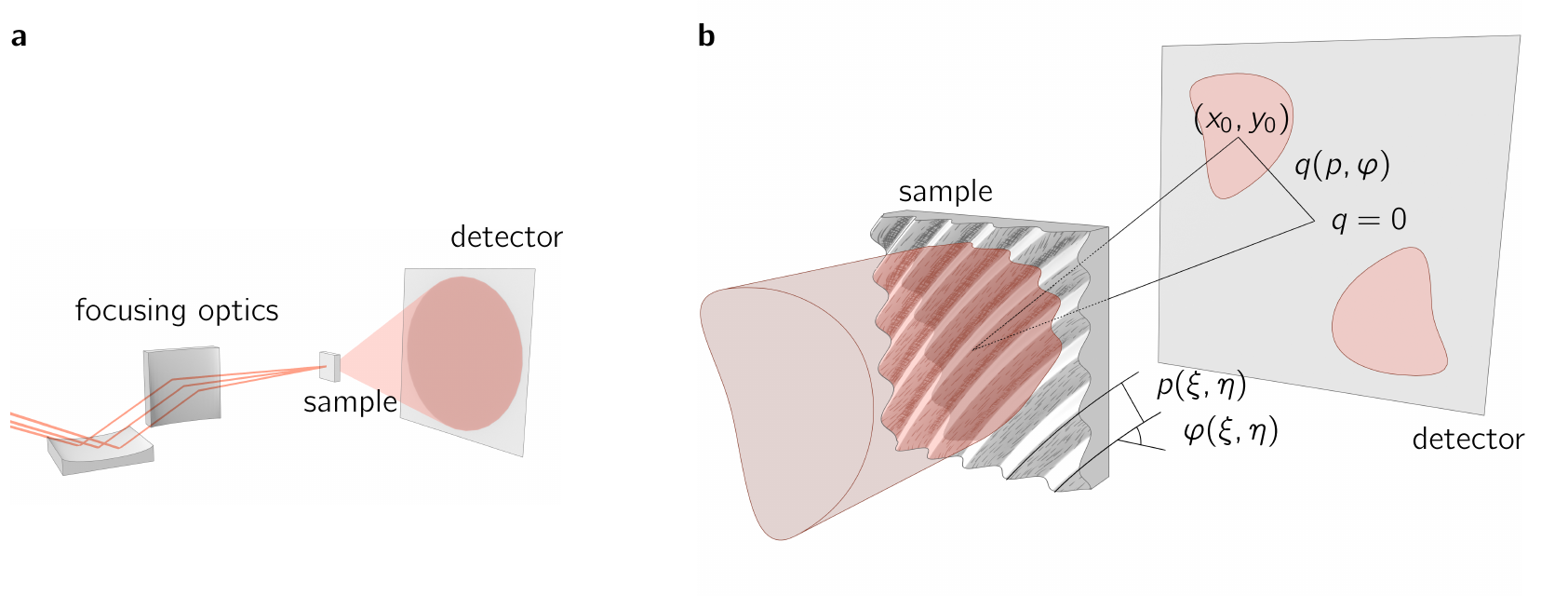}
    \caption{\textbf{Experiment geometry and sample design.}
    \textbf{a)} An optical system focuses the incoming beam (red lines) to a micrometer-sized spot on the sample.
    A 2D pixelated detector records the scattered radiation (red cone).
    \textbf{b)} The sample bears a suitably tailored, continuously varying grating with local periodicity $p(\xi, \eta)$ and local orientation angle $\varphi(\xi, \eta)$, where $\xi$ and $\eta$ are the coordinates in the sample plane.
    Incident light is diffracted away from the undeflected beam ($q=0$) with a momentum transfer of $\pm q(p, \varphi)$ according to the local grating parameters.}
    \label{fig:experiment}
\end{figure}

We perform an FEL diffraction experiment where the scattered radiation is detected as a function of momentum transfer $\vec{q}$ on a 2D pixelated detector, as shown in \cref{fig:experiment}a.
Note that this geometry also includes standard spectroscopy of the photon beam, where the beam at a selected momentum transfer (such as $q=0$) is detected as a function of wavelength.
In material and life sciences, thin membranes of \SiN, Si or polymers are commonly used to administer samples to the FEL beam.
We equip these membranes with a grating structure, that gives rise to an \emph{additional} scattering signal on the detector \cite{Schneider2016}.
The key idea of this work is to design the gratings such that each point on the sample surface diffracts the incoming light to a separate position on the detector while preserving the spatial relationship of the originating sample points.
In other words, the grating maps an image of the spatial photon distribution on the sample to a selectable detector region.
This is achieved by the following sinusoidal transmission function:
\begin{align}
    \centering
    t(\xi, \eta) &= \frac{1}{2} + \frac{1}{2}\cos(2\pi f(\xi, \eta)(\xi\cos(\varphi(\xi, \eta)) + \eta\sin(\varphi(\xi, \eta))))
    \label{eq:grating_transmission}
\end{align}
Here, $\xi$ and $\eta$ are spatial coordinates in the sample plane, while $f=1/p$ and $\varphi$ are the local spatial grating frequency and orientation angle, respectively.
They are given by the relations
\begin{align}
    \centering
    \varphi(\xi, \eta) &= \arctan((y_0 + m\eta) / (x_0 + m\xi))
    \label{eq:local_phi}\\
    f(\xi, \eta) &= \sin\left(\arctan(\sqrt{(x_0 + m\xi)^2 + (y_0 + m\eta)^2} / z)\right) / \lambda
    \label{eq:local_freq}
\end{align}
Here, $z$ is the distance from the sample to the detector, $(x_0, y_0)$ are the coordinates in the detector plane around which the intensity map is designed to be centered and $\lambda$ is the photon wavelength.
The dimensionless factor $m$ defines the magnification of the mapped intensity in the detector plane relative to the sample size.
We note that \cref{eq:grating_transmission,eq:local_phi,eq:local_freq} always constitute segments of Fresnel zone plates. 
One can alternatively obtain hyperbolic zone plate segments by adding a constant offset of $\varphi_0=\pi/2$ to the orientation angle $\varphi(\xi,\eta)$ in \cref{eq:grating_transmission}.
All gratings in the present work are of the hyperbolic kind, which offer better spatial resolution in the diffracted fluence maps (see Supplementary Notes). 

\begin{figure}
    \centering
    \includegraphics[width=170mm]{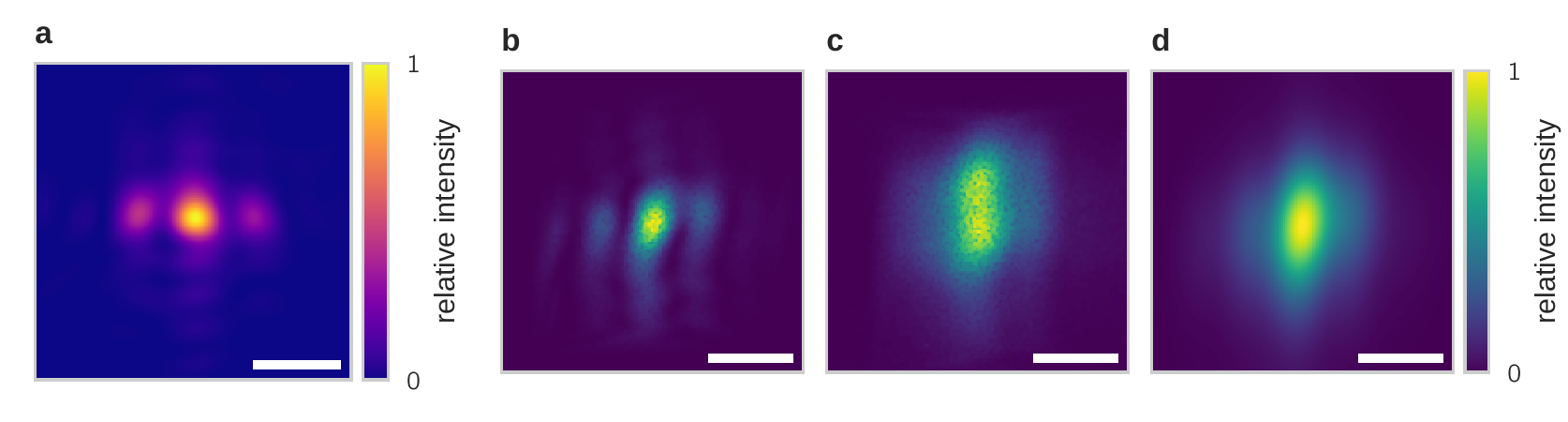}
    \caption{\textbf{Single- and multi-shot fluence distributions.}
    \textbf{a)} Single-shot wave-front sensor measurement of the photon distribution in the sample plane.
    \textbf{b)} Spatial fluence distribution of a destructive single-shot, measured by a curved grating sample.
    \textbf{c)} Accumulated fluence distribution of \num{3000} attenuated shots on the same grating.
    \textbf{d)} Single-shot diffraction pattern broadened by an asymmetric Gaussian distribution to simulate the multi-shot scattering pattern.
    Scale bars correspond to \SI{10}{\micro m}.
    For the entire series of \num{36} pairs of multi-shot and subsequent single-shot measurements, see Supplementary Video 2.
    }
    \label{fig:pointing}
\end{figure}

We present a spatial fluence distribution obtained in this fashion for a single FEL pulse in \cref{fig:pointing}b.
The map reveals a complex focal spot with a bright central area and several side lobes of considerable intensity.
This is in very good agreement with the independently obtained data for a different single-shot via a wave-front sensor measurement, shown in \cref{fig:pointing}a.
Small deviations from the actual photon distribution -- as measured by the wave-front sensor -- find their origin chiefly in the sample's non-negligible spatial extent:
The diffracted fluence map emanates from an extended source \emph{area}, as opposed to a singular point of origin.
Consequentially, each point in the diffracted fluence map is offset by its respective sample coordinate $(\xi, \eta)$, leading to a slight trapezoid distortion.

The focal spot's detailed structure is lost in the multi-shot fluence maps, as can be seen in \cref{fig:pointing}c.
Examining a series of \num{40} single-shot fluence maps (see Supplementary Notes and Supplementary Video 1) shows that the absolute position in the transverse direction to the optical axis varies from shot to shot, while the shape of the intensity within the focal spot does not change substantially.
Hence, the smearing out of this internal structure is due to the shot-to-shot pointing instability commonly encountered at FEL sources.
Such spatial jitter presents a major obstacle in the performance and interpretation of high-fluence, single-shot experiments, especially when inhomogeneous samples are investigated.
Given this situation, it is obvious that the interpretation of fluence-dependent phenomena is prone to large systematic errors, if the fluence on the sample has to be inferred from measurements over many shots or even from the integral pulse energy, as visualized in \cref{fig:error_schematic}.
In order to quantify the spatial jitter, we fit a blurred version of the single-shot fluence map to the multi-shot map (\cref{fig:pointing}d).
The blurring is achieved by a Gaussian function with separate standard deviations in horizontal and vertical direction as fit parameters.
We analyze the whole series of multi- and single-shot images and find the spatial jitter of the FEL beam on our sample to be \SI{4.2\pm0.4}{\micro m} and \SI{8.8\pm0.5}{\micro m} (full width half maximum) in horizontal and vertical direction, respectively.
It is obvious, that -- in addition to mapping the internal structure of the focus -- the access to the spatial jitter of the focus on the sample on a single-shot basis is extremely valuable.
This applies particularly to laterally inhomogeneous samples with spatially varying material composition, including particles sparsely dispersed on a membrane.

We note that spectral jitter, i.e.\ a shot-to-shot change of the photon wavelength, has a slightly different influence on the measured fluence maps.
Here, a centrosymmetric scaling of the diffraction pattern is to be expected due to the wavelength dependency of the scattering angles.
This effect is negligible for seeded FELs such as FERMI, but might become noticeable at FELs employing the self-amplified spontaneous emission (SASE) scheme, e.g.\ at LCLS or FLASH.

\begin{figure}
    \centering
    \includegraphics[width=170mm]{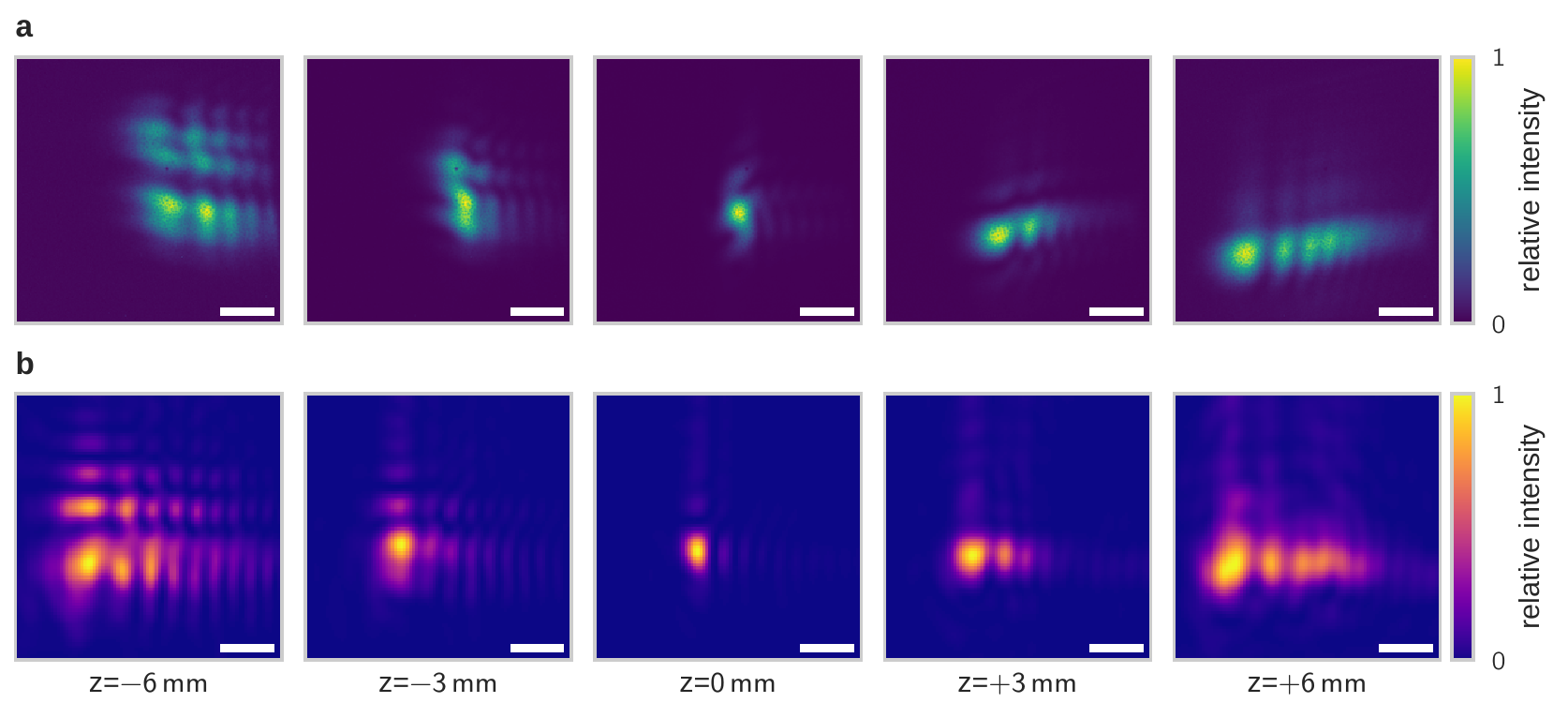}
    \caption{\textbf{Single-shot spatial fluence distributions along the beam propagation axis.}
    \textbf{a)} Positive grating diffraction order.
    \textbf{b)} Spatial fluence distribution calculated from wave-front sensor measurement.
    The position $z$ of the sample plane is given relative to the nominal focus position.
    Scale bars correspond to \SI{10}{\micro m}.
    For the entire caustic scan series, ranging from \ensuremath{z=\text{\SIrange{-9}{+9}{mm}}}, see Supplementary Video 2.
    }
    \label{fig:caustic}
\end{figure}

The spatial fluence distribution also changes substantially within a few millimeters along the beam propagation axis.
This is due to the finite size of the optical elements which act as limiting apertures and introduce diffraction artifacts into the beam.
The tailored grating integrated into the sample allows us to directly monitor this change for movement along the optical axis, as shown for selected positions along the beam propagation axis in \cref{fig:caustic}a (for the entire series see Supplementary Video 2).
Note that the images in \cref{fig:caustic} are from a separate experiment and thus are not expected to show the same spatial fluence distribution seen in \cref{fig:pointing}.
For all positions along the beam axis, we find excellent agreement of the fluence distribution's fine structure between our grating measurement and the results calculated from wave-front sensor data (\cref{fig:caustic}b).
However, the distribution's overall shape is sheared slightly stronger than what is to be expected from considering the grating's spatial extent.
This is an indication that -- for this particular experiment -- the conditions for the Fraunhofer approximation are not fully satisfied.
Potential distortions of the fluence map are discussed in the Supplementary Notes.
If required, the shearing could be corrected on this basis.

The fact that the spatial fluence distribution on the sample varies substantially for changes in the sample position on the order of one millimeter along the optical axis, again illustrates the potential for uncertainty when fluences on the sample are estimated from integral pulse energy measurements.

Our fluence-mapping approach is a unique tool for true \textit{in-situ}, single-shot-capable monitoring of the fine structure of the sample illumination in transmission-type scattering experiments.
It is, to our knowledge, the only approach that allows for a simultaneous, i.e.\ non-invasive, mapping of the fluence distribution together with a scattering signal of interest.
The approach provides an instantaneous online signal, which can be interpreted without any further computation, and can thus be used as instant feedback to align the upstream optical system.
In the study of fluence-dependent phenomena, it provides crucial information for the correct interpretation of the data.
The position and magnification of the photon-fluence map on the detector is, within the limits of the particular manufacturing process, freely selectable.
This makes our fluence mapping approach compatible with a large variety of experiments and samples.
Given these features, we expect this approach to become a valuable tool in particular for the study of non-linear light--matter interaction in the XUV and X-ray regime.

\section*{Methods}
We perform small-angle scattering in transmission geometry using the DiProI end-station at the FERMI@Elettra FEL source, where focusing is provided via a bendable Kirkpatrik--Baetz optics \cite{Raimondi2013}.
A Princeton Instruments PI-MTE in-vacuum charge-coupled device (CCD) camera (\num{2048x2048} pixel, \SI{13.5}{\micro m} edge length) detects the scattered radiation \SI{50}{mm} downstream of the sample with a cross-shaped beam-stop blocking the intense radiation in the forward direction and from potential membrane edge scattering.
The incident X-ray pulses are tuned to a wavelength of \SI{20.8}{nm} (\SI{59.6}{eV}) with single-shot pulse energies ranging from \SIrange{0.5}{60}{\micro J}.
For accumulating measurements at a repetition rate of \SI{10}{Hz}, solid-state filters and a gas absorber reduce the pulse energy to \SIrange{20}{80}{nJ} .
The samples consist of \SiN membranes of \SI{30}{nm} thickness with \SIrange{30}{200}{\micro m} edge length.
For the purpose of other experiments, a \SI{20}{nm} Co layer is deposited on the membranes by DC magnetron sputtering and capped with \SI{20}{nm} \SiN.
In the experiments reported here, this Co sample layer was in a uniform magnetic state and has no structural feature in the momentum transfer range sampled by the CCD.
We use a focused $\text{Ga}^+$ ion beam (FIB) to mill the gratings directly into the \SiN ~membrane using \SI{30}{kV} acceleration voltage and \SI{93}{pA} beam current.
For the milling process, the gratings are generated on a grid of \num{3500x3500} points with \SIparam{x_0=y_0}{4.1}{mm}, \SIparam{\lambda}{20.8}{nm}, \SIparam{z}{50}{mm} and $m=\text{\num{58}}$.
In this particular case, milling the \SI{35x35}{\micro m} grating with a nominal topographic amplitude of \SI{2}{nm} takes \SI{60}{s}.
A different sample is used for the caustic scan along the beam propagation axis.
Here, a bare \SI{100x100}{\micro m} \SiN membrane of \SI{30}{nm} thickness is patterned with a \SI{50}{\micro m} grating.
The remaining grating parameters are unchanged, except for \SIparam{x_0=y_0}{3.1}{mm} and $m=\text{\num{19}}$.
The distance from sample to detector in this case is \SI{100}{mm}.

\section*{Author Contributions}
MS, CMG, BP and SE conceived the experiment.
MS and CMG prepared the samples.
MS, CMG, BP, FC and EP conducted the scattering experiments.
MM, MZ, NM and LR performed the wave-front sensor measurements.
SE supervised the project.
MS and SE wrote the manuscript with input from all authors.

\section*{Competing financial interests}
The authors declare no competing financial interests.
\printbibliography

\end{document}